# Debootstrapping without Archeology: Stacked Implementations in Camlboot

Nathanaëlle Courant[a], Julien Lepiller[b], and Gabriel Scherer[a]

a   INRIA, France
b   Yale University, United States

**Abstract**
**Context**   It is common for programming languages that their reference implementation is implemented in the language itself. This requires a "bootstrap binary": the executable form of a previous version of the implementation is provided along with the sources, to be able to run the implementation itself.
   Those bootstrap binaries are opaque; they could contain bugs, or even malicious changes that could reproduce themselves when running the source version of the language implementation – this is called the "trusting trust attack". A collective project called Bootstrappable was launched in 2016 to remove bootstrap binaries, providing alternative build paths that do not rely on opaque binaries.
**Inquiry**   Debootstrapping generally combines a mix of two approaches. The "archaeological" approach works by locating old versions of systems, or legacy alternative implementations, that do not need the bootstrap binaries, and by preserving or restoring the ability to run them. The "tailored" approach re-implements a new, non-bootstrapped implementation of the system to debootstrap. Currently, the "tailored" approach is dominant for low-level system components (C, coreutils), and the "archaeological" approach is dominant among the few higher-level languages that were debootstrapped.
**Approach**   We advocate for the benefits of "tailored" debootstrapping implementations of high-level languages. The new implementation needs not be production-ready: only to be able to run the reference implementation correctly. We argue that this is feasible , with several side benefits besides debootstrapping.
**Knowledge**   We propose a specific design of composing/stacking several implementations: a reference interpreter for the language of interest, implemented in a small subset of the language, and a compiler for this small subset (in another language). Developing a reference interpreter is valuable independently of debootstrapping: it may help clarify the language semantics, and can be reused for other purposes such as differential testing of the other implementations.
**Grounding**   We present Camlboot, our project to debootstrap the OCaml compiler, version 4.07. Once we converged on this final design, the last version of Camlboot took about two human-months of implementation effort, demonstrating feasibility. Using diverse double-compilation, we were able to prove the absence of trusting trust attack in the existing bootstrap binaries of the standard OCaml implementation.
**Importance**   To our knowledge, this document is the first scholarly discussion of "tailored" debootstrapping for high-level programming languages. Debootstrapping recently grew an active community of free software contributors, but so far the interactions with the programming language research community have been minimal. We share our experience on Camlboot, trying to highlight aspects that are of interest to other language designers and implementors; we hope to foster stronger ties between the Bootstrappable project and relevant academic communities. In particular, the debootstrapping experience has been an interesting reflection on language design and implementation..

**ACM CCS 2012**
- **Software and its engineering** → Compilers; Interpreters;

## The Art, Science, and Engineering of Programming

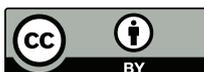

**Debootstrapping without Archeology: Stacked Implementations in Camlboot**

## 1 Introduction

### 1.1 Bootstrapped language implementations

Programming language designers and implementors devote themselves to creating the best possible programming language, and work hard on turning their language into usable software. What language would they use to implement their ideas? Their own programming language, of course! OCaml is implemented in OCaml, Haskell in Haskell, SML in SML, Scala in Scala, Rust in Rust, the list goes on and on. (For a criticism of the "language implementors should eat their own dogfood" trope, see Laurence Tratt's The Bootstrapped Compiler and the Damage Done.)

But then, newcomers ask, how can you run an OCaml program, if you need an OCaml implementation to run the OCaml implementation? This chicken-and-egg problem is solved by a *bootstrap*: along with the sources of the compiler, we carry a compiled form of the compiler (typically a slightly older version), which is used to build the compiler itself from sources, and we update this *bootstrap* binary from time to time. In the old times, a first implementation was written in some pre-existing programming language, but we replaced it with this bootstrap binary, which lets us write our compiler code in the designer's favorite programming language, their own.

The problem with this approach is that now we have to carry around this binary blob, version-control it, and we have to *trust* it to correctly compile our programming language. It is *opaque*: neither its content nor the nature of its changes in the repository can be inspected in practice. What if the bootstrap binary has a bug, could this bug somehow reproduce itself within the compiler compiled from the source? What if someone malicious was to intentionally insert malicious logic in the bootstrap binary that would compile the source compiler in such a way that it would insert a backdoor in compiled programs, and survive in later bootstrap binaries?

Bugs surviving through bootstrap binaries occur in practice, although rarely.[1] The idea of malicious backdoors hidden in bootstrap binaries is called a "trusting trust attack", popularized by Ken Thompson in 1983 [8]. Proof-of-concepts have been implemented, but we do not know if the attack has ever been used in practice.

### 1.2 Reproducibility and non-opaque build paths

The free software community has made a lot of effort to produce operating systems with reproducible/deterministic builds: building the same source package on different systems should produce the same binary (ideally bit-for-bit identical to ease comparison). This has been a constant effort in some communities, and was recently the main focus of the Reproducible Builds project started in 2013, which strives to make entire package distributions bit-for-bit reproducible.

---

[1] We heard of two such bugs from personal communication with programming-language implementers, one in a minimal Scheme and one in an early version of Caml.

Nathanaëlle Courant, Julien Lepiller, and Gabriel Scherer

Reproducible builds have obvious practical benefits: if you distribute binary packages and people don't know if they can trust you, they can compile the software independently and compare the hashes of their binaries with yours, to know for sure that you are distributing the right binaries.

However, even when the software is "reproducible" in this sense, this is achieved in most cases by building using an identical build environment provided by the untrusted provider. If the environment itself contains a version of the software you are trying to rebuild, how can you trust your own results? What if the build environment is corrupted? You would reproducibly get the same compromised result.

The Bootstrappable Builds project was started in 2016 to solve this issue, trying to eliminate opaque binaries from build paths, aiming for a more trustable operating system that can be built from nothing else than source code,[2] first in assembly language, then in increasingly higher-level programming languages.

The Bootstrappable effort is not yet complete, with some binary *seeds* (build path dependencies) remaining. For example, the more promising build paths for a large package ecosystem currently require build artifacts of a Scheme implementation as part of the seeds. Regular progress is being made to remove those binary seeds.[3]

Bootstrap binaries inside language implementations are the major source of opaque seeds in common build paths, so this problem should be of particular interest to the community of programming-language implementers. In fact, the feasibility of removing binary bootstraps is also related to language design aspects: this may be a new concern that language designers should consider.

Removing bootstrap binaries is important in practice because it reduces this build-environment dependency to source packages. They may still be the target of unintentional bugs or targeted attacks, but could be more effectively verified by humans or tools. This is also a better practice for system engineering and maintenance: it makes dependencies better-specified and more general (not "this binary" but "a tool satisfying <this interface>"), which eases re-implementation, cross-validation, and maintenance as the build environment evolves.

**A note on terminology**   The terminology around *bootstrap* is confusing and confused, with different and incompatible usage between some language-implementation communities and some operating-system and package-distribution communities. Many compiler authors call *to bootstrap* the act of building their language implementation using an executable version of itself – a *bootstrap binary*. In contrast, many operating-system people, in particular the bootstrappable-builds community, call *to bootstrap* the fact of building a toolchain from arbitrary build dependencies (the *bootstrap seeds*),

---

[2] "Source code" is loosely defined as the documents that are designed for human understanding of the program; typically they are the human-written vs. computer-generated artifacts, but the frontier is very blurry. Here "binary" means non-source. Different definitions on what is and is not part of source code result in weaker or stronger conditions for an implementation to be free from non-source bootstrap artifacts.

[3] See for example this description of the build path of GNU Guix, the software distribution of choice of the Bootstrappable people, described in 2013 by Ludovic Courtès [1].

which preferably do not include binaries corresponding to programs of the toolchain itself.

To avoid any confusion, we will use a *bootstrap binary* to mean the self-compilation build artifacts, and *to debootstrap* to mean the act of removing our dependency on bootstrap binaries, to allow a build from sources. A *bootstrapped* compiler uses a bootstrap binary, a *non-bootstrapped* or *debootstrapped* compiler does not.

In this paper, we report on our experience producing camlboot, a *debootstrapped* implementation of the OCaml programming language.[4]

### 1.3 Archaeology or tailored implementations?

The low-level systems debootstrapped by the Bootstrappable project have typically used tailored new implementations; a key component since 2016 is Jan Nieuwenhuizen's GNU Mes [5], a project to provide a Scheme interpreter (in C) and a small C compiler mescc (in Scheme) that can host themselves. Mes can be used to deboostrap TinyCC (tcc), a minimal C compiler, and from there build more standard C compilers.

On the other hand, higher-level components are typically debootstrapped using an archaeological approach, locating older versions that did not rely on a bootstrap binary, or legacy alternative implementations. Let us quote the beginning of the description of the Bootstrappable page on Java, to give an idea of the work involved:

> In Guix the Java bootstrap begins with Jikes, a Java compiler written in C++. We use it to build a simple version of GNU Classpath, the Java standard library. We chose version 0.93 because it is the last version that can be built with Jikes. With Jikes and this version of GNU Classpath we can build JamVM, a Java Virtual Machine. We build version 1.5.1 because it is the last version of JamVM that works with a version of GNU classpath that does not require ECJ. These three packages make up the bootstrap JDK.
>
> This is sufficient to build an older version of Ant, […]

In this work, we propose to use "tailored" implementations for debootstrapping purpose: software specifically designed for bootstrapping (adapted from existing projects or written from scratch), striving to remain as simple as possible.

This is more robust to future evolutions of our computing systems (legacy software in the build path may have trouble adapting to, say, RISC-V machines), and can reduce the overall complexity and build time of the build path.

### 1.4 Key metric: human work required to debootstrap

If we assume unlimited work resources, tailored debootstrapping is a trivial problem: just port your programming-language implementation to another language that has already been debootstrapped (for example C). But this is a massive effort that may never happen in practice – especially as you have to first convince your language implementors to work in a different programming language.

---

[4] https://github.com/Ekdohibs/camlboot, visited on 2022-02-02.

Debootstrapping becomes interesting once you consider human effort. Can you debootstrap your language implementation in a *reasonable* amount of work?

The idea is to build a *naive* implementation of your programming language. It can be slow, does not necessarily support all language features, its design would not necessarily scale to a full implementation. It can be produced with a much smaller effort and suffices to build your compiler.

Language design also matters. In particular, we relied heavily on OCaml's type-erasure property; a language has the *type-erasure* property if its runtime semantics can be defined independently of its typing derivation, that is if the untyped language already has a well-defined operational behavior. (Implicit casts/conversions may preserve this property, but stronger form of implicits such as type-classes do not.) This property lets you implement an interpreter without type-checking the programs first, saving the substantial effort of implementing a type-checker.

camlboot is the result of several design iterations, but the latest version was implemented from scratch in two human-months of work.

### 1.5 Diverse double-compilation

If we cut corners to build a *bad* debootstrapped implementation of a language, the maintainers of the better, bootstrapped implementations are not going replace their bootstrap binaries with our implementation. Have we gained anything?

Diverse double-compilation (DDC) [10, 11] is a technique proposed by David A. Wheeler in 2005 to use an alternative implementation to gain trust in a bootstrap binary, proving the absence of trusting trust attack. (It needs reproducible/deterministic compiler builds.)

- First, we use the bootstrap binary to build the reference implementation from source, and we check that the resulting binary is identical to the bootstrap binary; let us call this binary the *bootstrapped binary*.
- Second, we use our debootstrapped implementation to build the reference implementation under test. The result, the *debootstrapped binary*, may be very different from the bootstrapped binary (different or no optimisation, etc., as our debootstrapped compiler produces worse code), but it should have the same semantics.
- Finally, we use the *debootstrapped binary* to compile the reference implementation again, getting a *final binary*.

The final binary was produced without ever using the bootstrap binary, using the compiler sources from the reference implementation. If it is bit-for-bit identical to the bootstrapped binary, then we have proved the absence of trusting trust attack; if it is not, there may be a malicious backdoor or a self-reproducing bug, but there may also be a reproducibility issue in the toolchain.

We have successfully performed diverse double-compilation using our debootstrapped implementation, and successfully checked that the bootstrap binary of the reference OCaml implementation, version 4.07 [4] (released in 2018 and typically attributed to Xavier Leroy, Damien Doligez, Alain Frisch, Jacques Garrigue, Didier Rémy, and Jérôme Vouillon), is free of trusting trust attack.

## 1.6 Debootstrapping the OCaml compiler

We used the following approach to debootstrap the OCaml compiler version 4.07.1:

- Implement an interpreter for OCaml, that we call interp. More precisely, it covers the (large) subset of the language used in the reference compiler implementation. interp is itself written in a *smaller* subset of OCaml that we intentionally kept small, and call MiniML. This took a few human-weeks.
- Implement a compiler for MiniML in Scheme, that we call minicomp. This also took a few human-weeks (for a much smaller set of features than our interpreter supports). The compiler targets the OCaml bytecode, which has a C interpreter.
- We can then use our interp interpreter to interpret the reference OCaml compiler to build itself, without needing a bootstrap binary.

One aspect to be careful about is the other build dependencies than the compiler itself, notably the lexer and parser generators. OCaml 4.07 uses a parser generator, ocamlyacc, written in C. The lexer generator, ocamllex, is written in a small fragment of the language; we extended MiniML to cover it. We still had to "debootstrap" the lexer generator by writing by hand a lexer that could lex ocamllex's own input grammar!

The two-stage process (a naive interpreter compiled by a naive compiler) introduces massive inefficiencies in the build chain: we are spending computer time to save human time. On a specific build, we measured that running ocamlopt.opt (the OCaml native compiler, compiled with itself) is 27500x faster than interpreting the same ocamlopt compiler using our interp interpreter compiled with minicomp. In Section 6.2 we discuss these performance gaps in detail. Yet the whole debootstrapping process runs in human-reasonable time, completing under four hours on a developer machine.

We believe that such a staged design may be generalizable to many other language implementations. First write an interpreter of your language in a simple subset of itself – this is not too much work, and maximizes developer familiarity. Then implement the second stage as a compiler, in another language – this is again doable if the chosen subset is simple enough, and avoids an explosion of inefficiencies.

## 1.7 Results

We were able to debootstrap the OCaml compiler, and check using diverse double-compilation the absence of trusting trust attack in the bootstrapped compiler for OCaml 4.07.1. Our final design took less than two human-months to implement. Re-running the debootstrap chain to build the reference compiler takes under four hours. (In comparison, one parallel build of the OCaml distribution took 1m42s.)

A build recipe for GNU Guix is provided,[5] ensuring that anyone in the future will be able to reproduce this result thanks to very precise dependency information on the whole operating system configuration.

As a side-result, we produced a reference interpreter for a large fragment of the OCaml programming language. We ensured that this interpreter is clearly written and maintainable on its own; it has explanatory value, and we believe that it will be reusable in many other research venues. For example, having a small and simple reference implementation could be useful for fuzzing-based differential testing of realistic language implementations.

---

[5] https://issues.guix.gnu.org/46806, visited on 2022-02-02.

As a take-away for other language communities, we would recommend debootstrapping all implementations (it is challenging, interesting and fun), insist on the practical benefits of preserving type-erasure, and encourage language maintainers to write simple reference interpreters of their real-world programming languages.

**1.8 Related work**

The Bootstrappable Projects website describes heroic efforts to debootstrap various parts of the free-software ecosystem. It relies on GNU Guix (described in 2013 by Ludovic Courtès [1]), an operating system distribution that tracks dependencies in a fine-grained way, providing a tree of packages that rely on as few *bootstrap seeds* (non-source forms of programs required to build the system) as possible. Debootstrapping a bootstrapped programming language means that its bootstrap binary can be removed from the build seeds.

Guix is itself implemented in GNU Guile [12], a nice implementation of Scheme attributed to Andy Wingo, Marius Vollmer, Mikael Djurfeldt, Ludovic Courtés, and Jim Blandy, that started in 1993. The Guix, Guile and Scheme community at large are strong contributors to the Bootstrappable effort. Scheme is a natural choice for debootstrapping, being arguably the simplest programming language in which you might want to write a programming language implementation.

It is interesting to look at how other languages have approached debootstrapping. A short summary is that very few of the higher-level languages that have been relying on a bootstrap for a long time have been debootstrapped, a few exceptions being C, Rust and Java (but not Scheme!). OCaml is joining a select group with our work.

For C, Rust and Java, the debootstrapped path can build the most recent version of major implementations at the time of writing: gcc 11, openjdk 16.0, rustc 1.56.

**Rust** The reference Rust compiler, rustc, has been debootstrapped by David Tolnay in 2019 [9], making Rust the most advanced programming language with a good debootstrapping story so far. The debootstrap relies on John Hodge's mrustc [2], a partial reimplementation of Rust in C++ started in 2016. mrustc does not build recent versions of rustc, but it can build version 1.39.0, released in November 2019. Then 1.39.0 can be used to build 1.40, and so on, until release 1.56. Building each rustc version takes about 3 hours on a good machine, suggesting that the full debootstrapped build would take one or two days.

One point of interest is that mrustc implements type-checking for Rust programs but not lifetime inference, thanks to the "lifetime-erasure" property that lifetimes do not affect the dynamic semantics of the program. This is the ownership analogue of the type-erasure property we rely on in our own work.

**Debootstrapping without Archeology: Stacked Implementations in Camlboot**

**Java**    OpenJDK has been debootstrapped by Ricardo Wurmus in 2017 [14] by finding an elaborate path through legacy implementations, as we discussed in Section 1.3

**Haskell**    GHC has not been successfully debootstrapped yet. The most elaborate attempt so far was made by Ricardo Wurmus in 2017 [13], and managed to run the nhc98 compiler using the hugs interpreter – implemented in C. Unfortunately nhc98 lacks many modern Haskell features required to build the current GHC codebase.

**Gcc**    gcc has been successfully debootstrapped from much simpler implementations. This required two components:
- a C implementation able to build gcc, or rather, the latest version of gcc implemented in C instead of C++, gcc 4.7 (released in 2014), that can then build newer versions of gcc. This is provided by the TinyCC compiler, itself running on top of Jan Nieuwenhuizen's GNU Mes.
- Building gcc, or any modern software, also requires various core utilities (patch, sed), GNU Make, and a working Bash implementation. Building those without a modern C compiler is hard! The Guix project was able to remove them from the bootstrapping base by relying on Gash [7], a 2018 Guile implementation of Bash and core utilities attributed to Timothy Sample and Jan Nieuwenhuizen.

**Scheme**    Guile is *almost* debootstrapped. Many necessary pieces are present in the Guile implementation (it has a Scheme interpreter, written in plain Scheme, that can interpret the Guile compiler, and a plain Scheme interpreter in C that can interpret the interpreter), and the Scheme interpreter of GNU mes seems close to be able to run guile. At the time we wrote the present article, there was an issue with the macro-expander, which needed a pre-expanded version of itself as a bootstrap binary, but this issue appears to have been solved in January 2022, just not yet integrated upstream in the Guile project, and the interface with lower-level parts of a debootstrapped toolchain may still require some work.

Racket had a macro-expander implemented in C until version 6 included (Racket 6.12 was released in 2018). Version 7 moved to a macro-expander implemented in Racket, which was great for maintainability but introduced a bootstrap.

**Down to assembly**    There is work ongoing to connect GNU Mes "up" with Guile, and also (mes-m2) to connect it "down" with the stageo project [6], a 2017 project by Jeremiah Orians to build a tower of x86 assemblers from a single 1KB binary.

**Summary**    We found that the approach of writing entirely new implementations for the purpose of debootstrapping is more rarely used – to our knowledge, the only project to have used it before our work is GNU Mes.

In the general case, creating a debootstrapped "build path" to a language implementation may require a mix of legacy implementations, bootstrap replay, and new alternative implementations. New implementations written for debootstrapping may in turn become legacy applications if they are not maintained actively – and become incompatible with newer versions of the reference implementation. The build paths themselves need to be maintained; they can typically be lengthened by adding extra steps to build more recent versions of the reference implementation (for Rust, for example), or shortened by finding alternative, simpler build paths (by improving mrustc or, in our case, interp and minicomp). There are interesting maintenance challenges, and only time will tell whether the Bootstrappable community can tackle them.

Debootstrapping build plans are currently evolving rapidly; for example the live-bootstrap project appears to have recently managed a build of bash and make without relying on Gash/Guile. The global picture may be much improved by the time you read this article. But each bootstrapped programming language, OCaml in our case, still needs to be debootstrapped independently from the rest of the software ecosystem.

**SqueakVM** Some programming language implementations share the commonality of being written in their own language, with a small fragment compiled to a pre-existing language. One notable example is the Squeak VM, described in 1997 by Dan Ingalls, Ted Kaehler, John Maloney, Scott Wallace, and Alan Kay [3], which was implemented in pure Smalltalk, with its bytecode interpreter written in a simple fragment of Smalltalk designed to be directly translatable to C. However, this does not suffice to debootstrap the language: the compilation from Smalltalk to its bytecode, which is a more subtle piece of software, remains implemented in full Smalltalk, and requires a pre-existing VM image to run.[6] If the Squeak project wanted to use their "simple fragment" for larger part of the implementations – ideally all of it – they would probably want to add higher-level features of Smalltalk to this fragment. This would require a more elaborate compiler from their mini-Smalltalk to C (or some other target language), and would be closer to our design.

## 2 The OCaml compiler implementation

OCaml has one reference implementation that evolves along with the language itself, and a few alternative implementations that typically reuse some parts of the reference – to compile to JavaScript, etc.

The reference implementation provides *two* compilers: the *bytecode compiler* ocamlc, and the *native compiler* ocamlopt. ocamlc produces executables in a custom portable bytecode format, to be executed by the bytecode interpreter ocamlrun. ocamlopt produces native binaries directly, which are more efficient (typically 2x-10x) but not portable, larger and longer to compile. Back in the days of Caml Light, the bytecode was famous for its efficiency (among the fastest functional programming languages, due to its clever support for curried application), but nowadays people exclusively use the native compiler in production. Executables produced by the bytecode or native compiler link to a *runtime*, which is a bunch of C code and some assembly that implement low-level features the language relies on, in particular garbage collection.

---

[6] There is a project to partially debootstrap Squeak: SqueakBootstrapper.

**Debootstrapping without Archeology: Stacked Implementations in Camlboot**

[Figure 1: T-diagram showing the bootstrapped build of OCaml 4.07, with compilers for ocamlyacc (C→M), ocamllex (ML→B using boot/ocamlc), lexer.mll (MLL→ML using boot/ocamllex), ocamlc.byte, ocamlopt.byte, ocamlopt.opt (all ML→B/M using boot/ocamlc), and ocamlrun (C→M using gcc).]

B: OCaml bytecode. M: machine code. C: C source code. Scm: Scheme source code.
ML: OCaml source code. MLL: OCaml lexer definition. MLY: OCaml parser definition.
o: OCaml compiler sources.   red s: bootstrap seeds.

■ **Figure 1** The usual, bootstrapped build of OCaml 4.07

ocamlc and ocamlopt are implemented in OCaml. The runtime, as well as ocamlrun, are implemented in C code. The OCaml 4.07 compilers also use a parser generator, ocamlyacc, implemented in C (but the parsers it generates have an OCaml interface), and a lexer generator, ocamllex, implemented in OCaml.

**Bootstrap** To bootstrap itself, the OCaml compiler is distributed with bytecode executables for ocamlc (the bytecode compiler) and ocamllex. They can be executed by ocamlrun, which is built from C. To build the compiler from sources they also need to generate an OCaml parser from the yacc grammar, using ocamlyacc built from C. Before building the compiler sources it first needs to build the OCaml standard library, distributed with the compiler and used within the compiler sources.

We present in Figure 1 a schematic view of the usual OCaml build plan, using tombstone diagrams (T-diagrams).[7] The T-shaped blocks correspond to compilers – the left part of the T is the source language, the right part the target language, and the bottom part the language that must be interpreted to execute the compiler. I-shaped blocks are interpreters; ocamlrun interprets all the bytecode programs (B), and the other programs use machine code (M) that is interpreted by the processor directly. The Y-shaped blocks correspond to code files processed by the compilers as input, output, or both.

---

[7] We made some minor renamings and simplifications to the schema to keep it readable. For example, in the real build system lexer.mll is built using boot/lex, and ocamlopt.byte using boot/ocamlc.

For example, the ocamlyacc-related cluster describes how ocamlyacc is built as a native, machine program using gcc from its sources oyacc/*.c, and then used to turn the source grammar parser.mly – the OCaml grammar – into a .ml file included in the compiler codebase *.ml. Notice the placement of the gcc and ocamlyacc blocks, indicating that the output of gcc is the machine-code (M) program ocamlyacc.

The dotted arrows indicate aggregation: several different code files are included jointly, as inputs to a compiler or included in a larger set of code files. In this diagram, both ocamlyacc and ocamllex output files included in the *.ml source files. The full arrows indicate reuse: the same code files are passed as input to several compilers, or that the same interpreter runs several programs.

We use the (o :) marker to indicate sources from the standard OCaml distribution, and the red color and (s :) marker to indicate a bootstrap seed – a non-source program required to build the system. (Using markers in addition to color preserves the distinction on black&white medium.)

Note that oclex/lexer.mll represents the ocamllex lexer, used to lex the lexer-description input format (.mll file), not to be confused with the OCaml lexer lexer.mll.

## 3  A global view of our debootstrapped build path

Our debootstrapped build path is described in Figure 2. In addition to the previous conventions, we use the blue color and the (c :) symbol to denote our own Camlboot code, and the "mL" language (small "m") to denote ML sources that must stay within the MiniML fragment.

The build starts with both gcc and guile as seeds. Some parts of the build path of OCaml can be reused: we build ocamlrun, ocamlyacc and parser.ml the same way.

The first change is the construction of the sources of the compiler, more precisely the lexer, which was previously generated by boot/ocamllex.byte. We replace it by the version of ocamllex packaged with the OCaml sources, and compiled by minicomp. There is a difficulty here as well: these sources contain a lexer themselves, so we wrote a lexer by hand oclex/boot.ml, that was able to parse ocamllex's own lexer. We use it to generate a lexer for ocamllex, and use the result to generate ocamlc's lexer.[8]

---

[8] The description of the ocamllex.minibyte build in the figure is highly simplified. We combine our hand-written lexer with the sources of ocamllex, compile this with minicomp into a lex.boot.byte. Then we use lex.boot.minibyte to compile the reference lexer for ocamllex, lex/lexer.mll, and combine it with the other sources to produce our final ocamllex.minibyte. This two-build process, which is effectively implementing diverse double-compilation for the lexer, was necessary to ensure that the produced lexers are bit-for-bit" identical to the reference implementation.

**Debootstrapping without Archeology: Stacked Implementations in Camlboot**

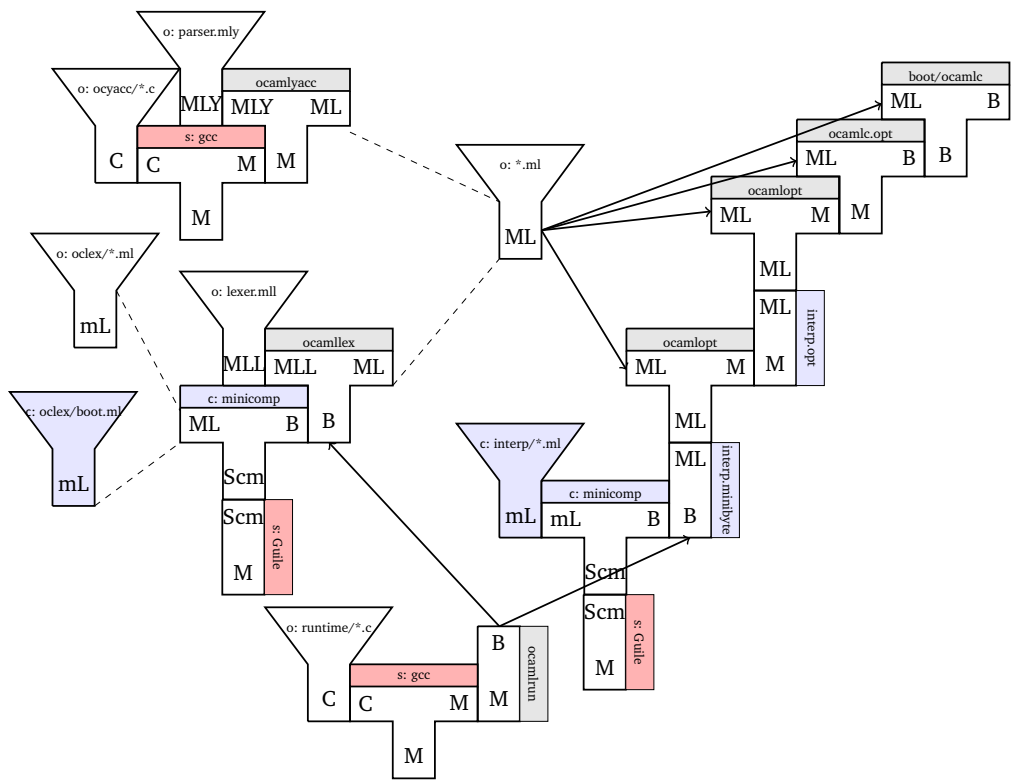

mL: miniML source code.
o: OCaml compiler sources.      red s: bootstrap seeds.      blue c: our Camlboot code.

■ **Figure 2** Our debootstrapped build of OCaml 4.07

Once we have the lexer and parser, we can compile our interpreter into interp.minibyte. We use it to interpret the ocamlopt compiler. For performance reasons, the first program we compile with this ocamlopt is our interpreter itself (see Section 6).[9] Then, with this natively compiled interpreter, we interpret ocamlopt again to compile the sources of ocamlc. With the resulting ocamlc.opt compiler, we can finally produce debootstrapped binaries to replace the OCaml bootstrap binaries. (We only represent the build of boot/ocamlc, but boot/ocamllex is rebuilt similarly.)

### 3.1 Defining the scope of MiniML

We did not define MiniML as a precise subset of OCaml when we started the project, but we refined it iteratively.

MiniML started with what many people would consider a canonical "toy ML": first-class functions, mutual recursion, variant/sum types and records. One notable limitation was on pattern-matching on variants/sums: the implementation would restrict to shallow patterns (no nesting of sum constructors) to avoid implementing pattern-matching compilation, a bulky feature.

---

[9] In the schema this is represented with two interpretation arrows: interp.minibyte interprets the sources of ocamlc, but it itself is interpreted by ocamlrun.

(In fact there was a more restricted version of MiniML in our first experiment, meant to be compiled to C instead of the OCaml bytecode, that would not support closures. We did not include these earlier implementations, which are gone from the final artifact, in our work estimates.)

First we wrote the interpreter; we tested it against the compiler codebase (we compile interp with the reference compiler for testing) and tried to support all necessary OCaml features with a simple, readable codebase.

Then we looked at the features we used in the interp codebase that were *not* supported by minicomp – outside the MiniML subset of the time. For each such feature, we had the choice to add it to MiniML and implement it in minicomp, or to rewrite the interpreter to stop using this feature, at the cost of uglier code.

In some cases features were motivated not directly by the interpreter codebase, but by their usage in the OCaml standard library for modules we wanted to use in the interpreter. In particular, we implemented functors to be able to use the Set and Map modules from the OCaml standard library. We want to be able to reuse the OCaml standard library in our interpreter code as much as reasonably possible, in the interest of maintainability. An alternative would be to implement our own specialized sets/maps data structures to not use functors (and in fact we did this in an intermediate version of our code). But while this makes debootstrapping slightly easier, it increases the maintenance burden on the project, duplicating code from the standard library, and it makes it "weird" as an OCaml project. We care very much about the fact that our interpreter can be taken as-is as a "good reference interpreter for OCaml", easy to maintain and usable by other people not interested in bootstrapping.

Moving from shallow patterns to full pattern-matching compilation was probably the most time-consuming and invasive addition; it required redesigning the minicomp compiler from one-pass (from AST to bytecode directly!) to a two-pass compiler, with pattern matching compiled into lower-level control-flow constructs. This took about a full week of work (using a not-too-naive compilation scheme), compared to a couple hours to implement full pattern-matching (naively) in the interpreter.

We believe that it is not a coincidence that our compiler and our interpreter consumed roughly the same amount of effort (about four human-weeks each). When deciding whether to extend the compiler or simplify the interpreter, we considered the effort involved on either end. Repeatedly going for the lowest estimated effort tended to evenly balance the time spent on both parts.

### 3.2 Extending the build path after OCaml 4.07

When this work started, 4.07.1 was the most recent version of OCaml, released in October 2018. The OCaml compiler follows a six-month release cycle, and several versions have been released since, the most recent at the time of writing being 4.12.

So far we have stuck with OCaml 4.07 rather than a more recent version, because OCaml 4.08 and later transitioned to Menhir, a parser generator itself written in OCaml, making the debootstrapping path more complex.

**Debootstrapping without Archeology: Stacked Implementations in Camlboot**

We leave a 4.12 build path as future work. There are two possible approaches, one is to try to build 4.12 from our debootstrapped 4.07 implementation, and the other is to update our debootstrapping tools to work with 4.12 directly. The first option corresponds to considering our implementations as "legacy", and we would rather explore the second one. A hybrid approach would be to build Menhir using our debootstrapped OCaml compiler (we checked that it builds with OCaml 4.07, even the recent versions used for the 4.12 parser), but upgrade our interpreter and compiler to support the rest of the 4.12 compiler implementation.

## 4 interp: Interpreting OCaml in MiniML

Our first software contribution is interp, an interpreter for OCaml written in MiniML, a reasonably small subset of OCaml. It takes 3000 lines of code, took about four human-weeks of work to write, and it covers all features of OCaml used in the standard library and the compiler, which is almost all the features of the language.[10]

### 4.1 A taste of the implementation

```
 1 type value = value_ Ptr.t
 2 and value_ =
 3   | Int of int
 4   | Int32 of int32
 5   | Int64 of int64
 6   | Nativeint of nativeint
 7   | Fun of arg_label * expression option * pattern * expression * env
 8   | Function of case list * env
 9   | String of bytes
10   | Float of float
11   | Tuple of value list
12   | Constructor of string * int * value option
13   | Prim of (value -> value)
14   | Fexpr of fexpr
15   | ModVal of mdl
16   | InChannel of in_channel
17   | OutChannel of out_channel
18   | Record of value ref SMap.t
19   | Lz of (unit -> value) ref
20   | Array of value array
21   | Fun_with_extra_args of value * value list * (arg_label * value) SMap.t
22   | Object of object_value
23 and fexpr = Location.t -> (arg_label * expression) list -> expression option
```

---

[10] Some runtime libraries not used inside the compiler, such as ephemerons and bigarrays, are not supported by our interpreter.

The interpreter is written to be reusable in other projects: it strives to be easy to read and maintain. To get a sense of it, we have shown its definition of OCaml "values".

Let us describe some technical details that show up in this type definition. We find them (mildly) interesting in that they show that writing an interpreter reveals interesting aspects of the programming-language semantics.

The value defined as value_ Ptr.t, where 'a Ptr.t describes a pointer to a value of type 'a that may be uninitialized, in which case it can be "backpatched" by assigning it exactly once. This mutable indirection is used to interpret OCaml recursive values.

The two other places where mutability occurs explicitly are Record **of** value **ref** SMap.t, as record fields may be mutable,[11] and Lz **of** (unit -> value) **ref**, representing lazy thunks that mutate themselves when forced.

The Fexpr **of** fexpr is used to represent certain compiler primitives, in particular short-circuiting operators a && b and a || b, whose operational semantics is given by rewriting/desugaring into other expressions, before their arguments are evaluated. In contrast, most primitives are defined by the Prim **of** (value -> value) case that take their arguments as values.

Here are a couple cases of the main evaluation function:

```
let rec eval_expr prims env expr =
  […]
  | Pexp_ifthenelse (e1, e2, e3) ->
    if is_true (eval_expr prims env e1)
    then eval_expr prims env e2
    else
      (match e3 with
       | None -> unit
       | Some e3 -> eval_expr prims env e3)
  | Pexp_try (e, cs) ->
    (try eval_expr prims env e
     with InternalException v ->
       (try eval_match prims env cs (Ok v)
        with Match_fail -> raise (InternalException v)))
```

The eval_expr function evaluates expressions to a result; prims is an environment to interpreter primitive/intrinsic names (it never changes, but its presence breaks a circularity between expression evaluation and primitive evaluation), env contains the value of term variables, and expr is the expression to evaluate.

In the Pexp_try case, evaluating a **try** .. **with** .. expression, we see that a program raising an exception is interpreted by raising an exception in the interpreter. We catch this exception, use eval_match to match its payload against the exception-handling clauses cs; if none of the exception-handling clauses matches the exception, then we propagate the exception to the ambient evaluation context.

---

[11] ML-family languages also have mutable references 'a **ref**. This is a derived construct in OCaml, defined as a record: **type** 'a **ref** = { **mutable** contents: 'a }.

**Debootstrapping without Archeology: Stacked Implementations in Camlboot**

Note: interp does not embed a parser for the OCaml surface language; it works directly from a parse tree as produced by the compiler parser. This is a natural choice for a reference evaluator, using the compiler as a library to parse in a compatible way, but it also works well in our debootstrapping scenario, where we can execute the compiler parser produced by ocamlyacc and interpret the resulting parse trees.

In the rest of this section, we go into the details of two technical issues we had to solve (among others). They support our impression that writing a reference interpreter for debootstrapping is full of insights on the language and its implementations.

### 4.2 Technical focus: module aliases

Writing a reference interpreter will confront yourself to hidden corners of your language design, whose operational semantics you may not be completely aware of. Language changes are proposed by people who are familiar with the reference implementation, a compiler, and may frame changes in terms that are natural from a compilation perspective, without working out the operational semantics precisely.

One such instance for us was "module aliases". OCaml has an ML module system, which supports nested modules and various forms of module bindings (including modules parametrized over modules). Files passed to the compiler are implicitly mapped into modules, so the file "foo.ml" will have its content put in a module Foo. However, these "toplevel module" names must be unique. The linker would reject two modules of the same name to avoid symbol name clashes. So a common idiom is to use long filenames for the toplevel modules, prefixed with the library name, and to provide a short "alias map" module that rebinds them to their shorter name. A library Lib may for example be implemented by two files Lib__Foo.ml and Lib__Bar.ml, with a helper file Lib.ml whose content is as follows:

```
1  module Foo = Lib__Foo
2  module Bar = Lib__Bar
```

This allows the user to write Lib.Foo, unaware of the long module name Lib__Foo. (In fact the long module names and the "alias map" module are typically generated by the build system, so they can be considered as implementation details for a poor implementation of namespacing.)

There is still a glitch with this approach: if Lib__Bar actually depends on Lib__Foo, can it refer to it through the nice name Lib.Foo, or should it use the ugly internal name Lib__Foo? Using Lib from Lib__Bar was originally rejected, as Lib itself mentions Lib__Bar and OCaml does not allow circular dependencies among files / toplevel modules. But users found it inconvenient, so the language semantics was changed slightly to allow it: the -no-alias-deps option, now widely used by default, decides that mentioning a module in the right-hand side of a module "aliasing" (rebinding), such as **module** Bar = Lib__Bar, does not introduce a strong dependency on this module: it is okay if the module is only present "later" in linking order. Of course, any actual usage of the Bar module introduces a strong dependency on its definition Lib__Bar, but merely mentioning Lib from Lib__Bar does not create a circular dependency.

When introduced, this feature was described from a compiler perspective: weak versus strong dependencies, linking order, etc. Giving its actual semantics as a reduction relation (big-step or small-step) turns out to be non-obvious.

Our first attempt was to model aliases using just variables. Evaluating a module name M, for example on the right-hand side of the binding **module** Foo = Lib__Foo, would just return "the free module name M", whether or not M had already been provided in the current environment. (We are evaluating modules in linking order, and a module is defined in the environment if it has already been evaluated.) Operations that require accessing the module structure, field access M.x for example, would then "force" module names into proper structures, failing if the module is not available in the current module environment. But this does not work in the presence of functors, which capture their definition-time module environment. If you pass "the free name M" as an argument to a functor, and the functor was defined in an environment where M is not available, then the functor will be unable to access the structure of its argument, while the caller of the functor may have had access to the definition of M.

In the end we gave up with trying to devise gentle semantics for -no-alias-deps. The toplevel modules of an OCaml program are big mutually recursive definitions, where modules can be defined *but not used* before they have been evaluated. More precisely, toplevel module names evaluate to a mutable cell that starts uninitialized, and gets "backpatched" into a complete structure when the module evaluation succeeds.

**Lesson 1.** *Maintaining a reference* interpreter *forces you to think about the semantic impact of compiler implementation changes. We recommend it for more informed language evolution.*

*Otherwise it is surprisingly easy to design, discuss, evaluate and integrate features described in terms of their compilation/elaboration semantics, without realizing that their direct operational semantics (as a rewrite relation, or in a reference interpreter) may be delicate.*

Note: Module aliases as a language feature lie on the boundary between the language constructs proper and the surrounding tooling (dependency management, linking semantics, etc.); we suspect that similar semantic difficulties may lie in this grey area for other languages as well.

### 4.3 Technical focus: interpreting ocamlc or ocamlopt?

The bytecode and native compilers, ocamlc and ocamlopt, share the same frontend (parser, type-checker, pattern-matching compilation, first pass of simplifications/optimizations). They have two different backends, with ocamlc having a much simpler backend. To give some numbers, the frontend is about 50K lines of code, the bytecode backend is 4K lines of code, and the native backend about 40K lines of code. Bytecode compilation is also noticeably faster than native compilation, typically twice faster.

When we set out to interpret the OCaml compiler, we thus started interpreting the bytecode compiler: less code to interpret, and the interpreted compiler would be faster to compile. Our attempt was thwarted by an unexpected coupling between the compiler and the language runtime.

**Debootstrapping without Archeology: Stacked Implementations in Camlboot**

In the object files (library archives or executables) produced by the bytecode compiler, there is not only the bytecode instructions for the program, but also various side-data such as tables of constant values and debug information. Most of these file formats are defined (in the compiler implementation) in a precise way, to be parsed by the bytecode interpreter `ocamlrun`: a header with a magic word and section tables in a precise binary format, bytecode instructions in a precise byte encoding, etc. But the constant tables and debug information are serialized using OCaml's built-in polymorphic pickling/marshaling operation (`input_value` and `output_value`), and inserted as-is in the bytecode object file.

This works fine with the reference implementation of the language, which uses the same implementation of (un)marshalling as `ocamlrun`: they are both using the implementation provided by the OCaml runtime system. But this does *not* work with our interpreter, which interprets the marshaling primitives differently: it calls the marshalling function on the interpreter representation of the value, which differs from the native representation, and thus we get different rules. If we use our interpreter to interpret the code of `ocamlc`, then the call to `output_value` in `ocamlc` source code (producing the bytecode artifact) will produce a serialized value in a different format, and `ocamlrun` will crash when trying to deserialize this part of the bytecode executable and consume it.

To work around this issue, we could have modified the implementation of `ocamlc` and `ocamlrun` to stop using OCaml's built-in marshaling functions, and instead use a precisely defined binary serialization format for constant tables and debug information. But this would be a non-trivial change from the reference implementation, and it is unclear that the upstream maintainers would have been willing to integrate it. We would like our debootstrapping process to be maintainable alongside the reference implementation, rather than requiring the application of nontrivial patches to get a debootstrappable compiler.

So we decided to give up interpreting the bytecode compiler `ocamlc`, and interpret the native compiler `ocamlopt` instead. The native compiler produces binaries in its standard system format (ELF on Linux), without any implementation-defined parts. We had to extend our interpreter with a sizeable amount of language features used in the `ocamlopt` codebase but not in `ocamlc`, notably OCaml objects. The backend uses class inheritance to factorize code over several architectures – for example, instruction selection is defined in a main class, that is inherited in each backend to implement architecture-specific refinements.

**Lesson 2.** *The implementation of your compiler may of course use various primitives of your language with implementation-defined behavior. But it is not a good idea to have its* output artifact *format be implementation-defined. They should be defined so that different implementations can easily achieve cross-compatibility.*

### 4.4 interp language coverage

We implemented OCaml features "on demand" by iteratively trying to run the compiler sources, which would fail on unsupported features. If we had been able to run the bytecode compiler, we would probably have left objects (used only in the native-compiler backend) unsupported, as interpreting them was a fair amount of work.

Here are the OCaml features that are currently missing from interp, and not used in the compiler codebase:

- lazy patterns (**lazy** *p* will force its scrutinee and match its value with *p*),
- a couple minor pattern features (polymorphic variant type patterns **#foo**, local module open in patterns M.(*p*),
- direct object expressions (**object** .. **end** outside a class declaration),
- and recursive modules.

Most of these would be fairly easy to add; recursive modules may be a more sizeable addition, but we already support recursive compilation units.

## 5 minicomp: Compiling MiniML to the OCaml bytecode

Our second contribution is minicomp, a compiler for MiniML to OCaml bytecode, written in Scheme (more specifically, guile). It is comparable in terms of complexity to interp, taking about 3300 lines of code and having taken about four human-weeks of work to write. The feature set is more restricted than for interp: it does not handle objects, classes, lazy values, first-class modules or format strings; functors are generative and compiled by defunctionalization, and type-based disambiguation of constructor names or record labels is unsupported as well.

For the frontend, minicomp uses the lalr-scm parser generator, and a handwritten lexer. In the backend, minicomp produces OCaml bytecode, so that:

- We can use OCaml runtime primitives, including in the generated lexer and parser.
- We get good integration with an efficient garbage collector.
- We get efficient support for closures and curried functions.

An early experiment compiled MiniML to C code linked with the OCaml runtime instead; it produced less efficient code, due to having to register every intermediate value as a GC root and having to check for exceptions at each call. In addition, it did not handle closures, which made programming inside its MiniML subset quite painful.

The compiler itself is divided into two passes: a first pass, *lowering*, simplifies input expressions by compiling pattern matching to lower-level constructs, handles function application (recognition of tail calls, primitive calls, reordering labeled arguments), and transforms constructors and records into blocks with a fixed integer tag. The second pass compiles these simpler expressions and outputs the result directly to a bytecode file (no intermediate representation of bytecode exists in the compiler), backpatching labels as necessary. Most constructions of the lowered expressions map directly to bytecode, and we only need to take care of the scope of variables (local, belonging to the environment of a closure or global), and to the size of the stack.

We include below a representative fragment of the compile-expr function. It takes as input an environment for locating the variables, the current size of the stack, and a lowered expression to compile. It compiles the input by directly writing bytecode opcodes to the output file, with functions such as bytecode-put-u32-le. The functions bytecode-emit-label and bytecode-emit-labref record the current position for future references to the label, and modify the output file (using seek) to overwrite any previous reference to the label with its actual position.

```
1  (define (compile-expr env stacksize expr)
2    (match expr
3      [...]
4      (('LTailApply e args)
5       (let* ((nargs (length args)))
6         (compile-args env stacksize args)
7         (bytecode-put-u32-le PUSH)
8         (compile-expr env (+ stacksize nargs) e)
9         (bytecode-put-u32-le APPTERM)
10        (bytecode-put-u32-le nargs)
11        (bytecode-put-u32-le (+ stacksize nargs))))
12     (('LIf e1 e2 e3)
13      (let* ((lab1 (newlabel))
14             (lab2 (newlabel)))
15        (compile-expr env stacksize e1)
16        (bytecode-put-u32-le BRANCHIFNOT)
17        (bytecode-emit-labref lab1)
18        (compile-expr env stacksize e2)
19        (bytecode-BRANCH-to lab2)
20        (bytecode-emit-label lab1)
21        (compile-expr env stacksize e3)
22        (bytecode-emit-label lab2)))
```

## 6 Compiling OCaml with our interpreter

Once we had all the pieces in place, we can debootstrap the OCaml compiler. The idea is to *compile* the programs ocamlc and ocamllex from the reference implementation using the reference ocamlopt native compiler, *interpreted* from source by our interp interpreter, itself *compiled* into OCaml bytecode by our minicomp compiler and then *interpreted* by the reference ocamlrun bytecode interpreter, Phew!

In this section, we adopt precise yet concise notations to talk about a particular way to run a particular implementation. For an OCaml program foo, we write foo.opt for the native binary produced by ocamlopt, foo.byte for the bytecode binary produced by ocamlc, and foo.minibyte for the bytecode produced by our naive minicomp compiler. "Running" any of those programs means either running the native code directly, or using ocamlrun to run the bytecode. We also write $f$ (foo) to talk about the action of running foo interpreted by an interpreter $f$. To reformulate the previous paragraph, this section is about running interp.minibyte(ocamlopt) to compile ocamlc into ocamlc.opt, and from there produce clean binaries to replace the bootstrapping copies of ocamlc.byte and ocamllex.byte.

■ **Table 1** Timing of our Three Build Plans

|               | First     | Optimized | Parallel |
|---------------|-----------|-----------|----------|
| ocamlrun      | 1m        | 1m        | 1m       |
| interp.minibyte | 2m      | 2m        | 2m       |
| interp.opt    | not built | 8h56m     | 2h02m    |
| stdlib.opt    | 4h40m     | 48m       | 23m      |
| ocamlc.opt    | 25h40m    | 4h08m     | 1h31m    |
| Total         | 30h23m    | 13h55m    | 3h59m    |

## 6.1 Build times

You will not be surprised to hear that naively interpreting a large, complex program is *slow*, and that running that interpreter as a naively compiled program becomes *really slow*. Therefore, we iterated over several build plans in order to make the experiment fast enough to be reproducible.

All our build plans start by building ocamlrun from the C sources of the OCaml compiler distribution, which takes around one minute, and then run minicomp to compile interp into interp.minibyte. This step takes around two minutes – in any case, these times are negligible compared to the time taken for the other steps. The complete timings for the build plans are summarized in Table 1. All times were measured on a machine equipped with an Intel(R) Core(TM) i7-8665U CPU @ 1.90GHz CPU (4 cores, 8 threads) and 16 GB of RAM.

**First build plan**   Our first complete debootstrapped build took the following steps:
1. Run interp.minibyte(ocamlopt) to compile the OCaml standard library, a dependency of the compiler sources.
2. Run interp.minibyte(ocamlopt) to compile ocamlc into ocamlc.opt.

At this point we were not completely finished, but we knew we could use ocamlc.opt to continue the debootstrap in reasonable time (building the whole compiler codebases from it takes a few minutes at most). However, the total running time was around 30 hours, which we needed to optimize.

**Improved build plan**   We improved the code generation of minicomp slightly, which gave us a 10-20% performance improvement in interp.minibyte, and we added an extra step: instead of using interp.minibyte(ocamlopt) to compile the standard library then ocamlc, we would use it to compile interp.opt first, and then use that faster interpreter for the remaining compilation steps. This improved build plan is as follows:
1. Run interp.minibyte(ocamlopt) to compile the interp into interp.opt.
2. Run interp.opt(ocamlopt) to compile the standard library.
3. Run interp.opt(ocamlopt) to compile ocamlc.

This extra step shrinks the total build time from around 30 hours to around 14 hours.

**Debootstrapping without Archeology: Stacked Implementations in Camlboot**

**Parallelized build plan**   None of the many implementations discussed here are doing any active effort to parallelize builds, but compiling a codebase in independent files/modules leads itself naturally to makefile-level parallelism. With parallel builds enabled, we got actually reasonable build times: the total build time shrank from around 14 hours to around 4 hours.

At this point it is reasonable to expect other people to reproduce the experiment, and check that our scripted diverse double-compilation confirms the absence of trusting trust attack in the reference bootstrap binaries.

## 6.2 Performance analysis

We compare compile times for the interp codebase, without any parallelism:
- with ocamlopt.opt: 1.7s
- with ocamlopt.byte: 5.8s (3.4x slower)
- with interp.opt(ocamlopt): 2h30mn (1551x slower than 5.8s with ocamlopt.byte)
- with interp.minibyte(ocamlopt): 13h (5.2x slower than 2h30 with interp.opt)

This suggests that the performance gap between interpretation and compilation is much larger than between bad compilation and better compilation.

On other (smaller) compile targets, we observed that
- interp.byte is around 2.5x slower than interp.opt, and
- interp.minibyte is in turn 2.2x slower than interp.byte.

It is interesting that the performance cost of bytecode compared to native code is similar to the performance cost of our "naive bytecode generation" compared to the decent optimization and code-generation work of ocamlc. This suggests that we could improve performance of our debootstrapping toolchain by improving the bytecode generated by minicomp. Unfortunately, the lack of convenient profiling tools for bytecode programs makes it hard to determine our main sources of inefficiencies in the produced bytecode.

**Implementing an interpreter in Scheme?**   Our approach implements both an interpreter and a compiler. Should we instead implement an interpreter directly in another, simpler programming language?

To evaluate this design alternative, we implemented a naive Scheme interpreter for a small fragment of OCaml – essentially MiniML.[12] Before doing performance measurements, we did not know whether it would turn out faster or slower than our compiled-interpreter approach:

- Both interpreters (in OCaml and in Scheme) are implemented in a similar, very naive "reference interpreter" style, without much care for performance; both could probably be optimized with some care.
- Our MiniML interpreter is compiled using our naive minicomp compiler, introducing inefficiencies compared to a production implementation. In contrast, our Scheme interpreter is run with Guile, which is a reasonable implementation of Scheme. Guile compiles its programs to a bytecode internally (just as our compiler does), and even provides a JIT – performs Just-in-Time code generation.
- On the other hand, using a dynamic language typically incurs a performance overhead, due to extra dynamic checks and less compact, more self-describing representations of data structure. Some implementations use advanced speculative optimizations to remove this overhead, but Guile is not that sophisticated yet.

---
[12] https://github.com/gasche/camlboot/tree/scheme-interpreter, visited on 2022-02-02.

On a small test program, we observed that
- interp.minibyte is 6.1x slower than interp.opt. (This is roughly consistent with our measurements interpreting the whole interp codebase.)
- guile interp.scm, our Scheme interpreter running with Guile, is 4.0x slower than interp.minibyte.

One could try to run the interpreter with another Scheme implementation, or switch to an implementation language without dynamic overhead, such as C. In any case, our conclusion is that our compiler+interpreter design is in fact sensible, performance-wise.

Note that performance was not the only guiding factor in our choice:
- Writing the interpreter in OCaml itself has value for OCaml programmers needing a reference interpreter.
- Targeting the OCaml bytecode lets us reuse the OCaml runtime system, which greatly simplifies the implementation of runtime primitives in our interpreter: we can let runtime primitives be implemented by themselves.

If we wanted to complete our Scheme interpreter to support a reasonable fragment of OCaml, we would have to either reimplement those runtime primitives in Scheme, or convert value representations to the native OCaml representation to use FFI bindings to the OCaml runtime. This may be more work than our approach of implementing a naive compiler to OCaml bytecode, and would again incur a performance overhead.

# 7 Conclusion

Debootstrapping a language implementation by writing new code is a highly rewarding adventure. Our approach was to write an interpreter in the language itself, staying inside a simple enough fragment that we could write a compiler for. Of course, you now also have the possibility of using OCaml to implement your interpreter.

As a side-result, we wrote a simple interpreter for a large subset of OCaml; this already forced us to think hard about the operational semantics of newer language features. We expect that it will find various unplanned use-cases in the future.

We checked the absence of trusting trust attack in the version 4.07 of the OCaml compiler. In the medium-term future, we want to debootstrap more recent versions of OCaml as well, and think about how to maintain our debootstrap toolchain to follow the evolution of the language and its reference implementation.

## A  Feedback from OCaml programmers on Guile for compiler implementation

*(One of our reviewers asked for more details on our experience using Guile for debootstrapping. We are not sure the subjective opinion of the authors has value; and we certainly suffer from a familiarity bias in favor of statically typed functional programming languages. But here it is.)*

We have good things to say about Scheme in general and Guile in particular, they are certainly nice languages. The usability of Guile is good, the tooling is solid, and we found nice libraries for our needs, in particular the lalr macro is very pleasant.

Our main source of frustration was the absence of static typing. Writing a compiler involves a lot of choices of data representation that create coupling between different parts of the program (typically: the generator of some part of the IR, the consumer of the same part of the IR); it happened time and time again that we would change one of these data representations, and have the code break in various places with errors that were always obvious (and fairly immediate thanks to test coverage) but also always more painful to understand and act upon than a static typing error.

Typed Racket would definitely have been an improvement for this — but there are less Racket users interested in debootstrapping than Guile users, so it sounds like a more risky choice. A systematic use of contracts may also have helped, and may be something to consider in the medium term for our project. We hope that eventually a Typed Guile will exist.

**Debootstrapping without Archeology: Stacked Implementations in Camlboot**

## About the authors

**Nathanaëlle Courant**
nathanaelle.courant@inria.fr
orcid: 0000-0002-8736-3060

**Julien Lepiller**
julien.lepiller@yale.edu
orcid: 0000-0003-2284-5488

**Gabriel Scherer**
gabriel.scherer@inria.fr
orcid: 0000-0003-1758-3938